\documentclass[fleqn,usenatbib]{mnras}
\usepackage[utf8]{inputenc}
\usepackage[T1]{fontenc}
\usepackage{graphicx}
\usepackage{amsmath}
\usepackage{amssymb}

\title[On the Internal Structure of Relativistic Jets]
{On the Internal Structure of Relativistic Jets with Zero Velocity Along the Axis}
\author[V.~S.~Beskin, F.~A.~Kniazev, and K.~Chatterjee]
{V.~S.~Beskin$^{1,2,3}$\thanks{E-mail: beskin@lpi.ru (VB)}, F.~A.~Kniazev$^{2}$, 
 and
K.~Chatterjee$^{4,5}$
\\
$^{1}$P.N.Lebedev Physical Institute, Leninsky prospekt 53, Moscow 119991, Russia \\
$^{2}$Moscow Institute of Physics and Technology (State Research University), Institutsky per.~9, Dolgoprudny 141700, Russia \\
$^{3}$National Research Center "Kurchatov Institute", Kurchatov sqr.~1, Moscow, 123182, Russia \\
$^{4}$Black Hole Initiative at Harvard University, 20 Garden Street, Cambridge, MA 02138, USA \\
$^{5}$Harvard-Smithsonian Center for Astrophysics, 60 Garden Street, Cambridge, MA 02138, USA \\
}

\begin{document}

\date{Accepted. Received; in original form}

\pagerange{\pageref{firstpage}--\pageref{lastpage}} \pubyear{2018}

\maketitle

\label{firstpage}

\begin{abstract}
The present work is devoted to the analysis of the internal structure of relativistic jets
under the condition that the velocity of the plasma flow at the jet axis vanishes. It is
shown that in spite of the seemingly fundamental difference in the formulation of the problem
at the axis, the key properties of the internal structure of such relativistic jets remain 
the same as for nonzero velocity along the axis. In both cases, at a sufficiently low ambient  
pressure, a dense core appears near the axis, the radius of which is close to the size of 
the light cylinder. 
\end{abstract}


\section{Introduction}

The significant progress of radio interferometry 
at long baselines makes it possible to directly explore the
internal structure of relativistic jets from active galactic 
nuclei~\citep[AGN;][]{Gabuzda, 2007ApJ...668L..27K, MOJAVEVIII, MOJAVE_XIII, MOJAVEXVI, Mertens, Zobnina} 
which are visible manifestations of their activity at an early stage 
of evolution~\citep{BBR84, Urry, Tche, Porth}. Such detailed observational studies allow us 
to test the numerous predictions of the theory 
of strongly magnetised outflows that have been developed since the 
1970s~\citep{L76, B76, Camenzind86, HN89, Camenzind90, Tomimatsu, Chiueh91, PP92, Appl92, Bogovalov92, BP93, Eichler93, LHAN99, BM00, BN06, Lyu09, BN09}. 
The main conclusions of these theoretical papers, discussed in several reviews and monographs~\citep{BBR84, Heyvaert96, Krolik, Camenzind07, MHD, Meier}, 
were later confirmed by numerical simulations of jets from accreting black holes~\citep{Ustyugova95, Ustyugova, McKinney06, Komissarov07, Romanova, Tche08, Porth_etal11, McKinney12, Kousheta}.

\begin{figure} 
\begin{center}
\includegraphics[width=0.95\columnwidth]{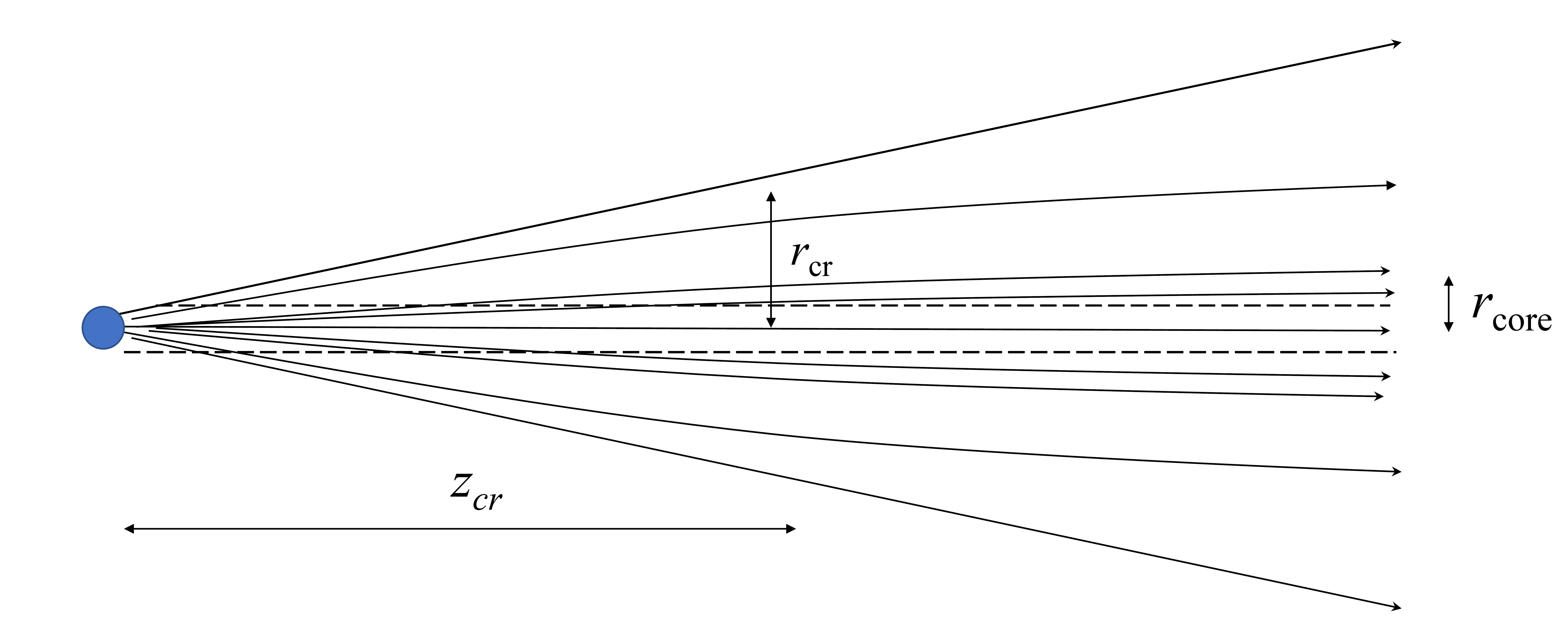}
  \end{center}
  \caption{The structure of the magnetic field in the model under consideration. 
  At a distances $z > z_{\rm cr}$ from the central engine,  when the transverse size of 
  the jet reaches the scale $\sim r_{\rm cr}$, in a conical flow (in which the 
  plasma density and the magnetic field weakly depend on the distance from the axis), 
  a denser central core begins to form. The light cylinder is shown by a dashed line.
  }
\label{fig0}
\end{figure}

One of these theoretical predictions repeatedly confirmed by numerical simulations
is the existence of a universal asymptotic behaviour for the Lorentz factor of an outflow
$\gamma = \varpi/R_{\rm L}$, where $\varpi$ is the distance from the rotation axis, 
and $R_{\rm L} = c/\Omega$ is the radius of the light cylinder  ($\Omega$ is the angular velocity of
the central engine). As another example, one can mention the presence of a central dense cylindrical 
core with the radius
\begin{equation}
r_{\rm core} = u_{\rm in} R_{\rm L},
\label{core}
\end{equation}
where $u_{\rm in}$ is the four-velocity of a flow along rotation axis. This result was first obtained 
analytically~\citep{Chiueh91, Eichler93, Bogovalov95, Bogovalov98, BM00, BN06, Lyu09, BN09} 
and later confirmed numerically~\citep{Komissarov07, Tche08, Porth_etal11}.
As shown in Figure~\ref{fig0}, this core is formed over long enough distances $z > z_{\rm cr}$ 
from the central engine when the transverse dimension of the jet $r_{\rm jet}$ becomes larger than 
$r_{\rm cr} = (u_{\rm in}\sigma_{\rm M})^{1/2} R_{\rm L}$. Accordingly, the poloidal magnetic field
at this distance $B_{\rm cr} = B_{\rm p}(z_{\rm cr})$ becomes equal to
\begin{equation}
B_{\rm cr} = \frac{B_{\rm L}}{\sigma_{\rm M}u_{\rm in}}.
\label{Bcr}
\end{equation}
Here $\sigma_{\rm M}$ is the Michel magnetisation parameter, 
and $B_{\rm L}$ is the magnetic field on the light cylinder near the origin (see formal definitions 
below).
It is necessary to emphasise that relation (\ref{core}) was also verified for non-relativistic 
flows, i.e. for $u_{\rm in} \ll c$~\citep{LHAN99, BoTsi99, TsiBo02, BN09}. 

We emphasise that, as was already known in the late 1990s, the internal structure
of relativistic jets is very sensitive to the behaviour of the Grad-Shafranov (GS) solution 
near the axis~\citep{Chiueh91, Eichler93, Bogovalov95, Lyub1997}. The difficulty of solving 
the GS equations in this region proved to be the stumbling block that did not allow us to
link together the various asymptotic solutions obtained. Only after the work by~\citet{BM00}
did it become clear that the central core exists only for sufficiently low ambient medium 
pressure $P_{\rm ext} < P_{\rm cr}$ (i.e., at sufficiently large distances from the central 
engine), where $P_{\rm cr} = B_{\rm cr}^2/8\pi$. For larger ambient pressures 
$P_{\rm ext} > P_{\rm cr}$ (i.e. at small distances $z < z_{\rm cr}$), the poloidal 
magnetic field remains practically constant within the whole jet. 

As a result, depending on the ambient pressure  $P_{\rm ext} < P_{\rm cr}$, the poloidal magnetic field 
$B_{\rm p}$ outside the central core has the form  
\begin{equation}
B_{\rm p} \propto \varpi^{-\alpha}
\label{3}
\end{equation} 
with $0 < \alpha < 1$. At the same time, however, the magnetic field in the core itself does
not differ significantly from the value $B_{\rm cr}$. In this case, the jet 
remains magnetically dominated till the distance from the origin when the external pressure drops to 
$P_{\rm ext} \approx P_{\rm eq} = B_{\rm eq}^2/8\pi$, where $B_{\rm eq} = \sigma_{\rm M}^{-2}B_{\rm L}$. 
At lower ambient pressures, the flow becomes particle-dominated.

Here, however, one important remark should be made. This model explicitly assumed 
that the flow velocity along the jet axis itself does not vanish. In fact, relation 
(\ref{core}) that, in the non-relativistic regime (i.e. $u_{\rm in} \rightarrow 0$),
the core radius \mbox{$r_{\rm core} \rightarrow 0$} as well. Thus, in the non-relativistic
regime, $r_{\rm core} \rightarrow 0$ when $u_{\rm in}$ tends to zero. Thus, the very existence of 
the central core is called into question.  Whether this result 
remains valid if the flow velocity vanishes on the jet axis has not been considered 
in detail up to now.

It must be said that the very assumption that the velocity on the jet axis is not equal 
to zero still had some grounds. It is based on the model of plasma generation in the
vacuum region near the black hole surface~\citep{BIP92, HO98, Ptitsyna, Crinquand}, which 
is equivalent to the so-called ``outer gap'' in the magnetosphere of radio pulsars. In 
this case, the value $u_{\rm in}$ arises as a natural boundary condition for the Grad-Shafranov
equation~\citep{BK00}, which ultimately leads to the existence of a central core.

On the other hand, there is also support for outflow models with zero velocity along the jet axis. 
For example, this occurs when the only mechanism of plasma acceleration is via 
electromagnetic forces (the Poynting vector flux on the jet axis is equal to zero). This 
point of view can also be supported by the pioneering work of~\citet{Tomimatsu}, who 
introduced the notion of a stagnation point, i.e. the region of the base of the flow 
where the velocity is zero. It was further shown that the hydrodynamical motion in a strongly 
magnetised flow is completely determined by the electric drift; the motion along the 
magnetic field lines can be neglected~\citep{Tche08, MHD}.
Despite the fact that this result concerns 
only the asymptotically far region $\varpi \gg R_{\rm L}$, it began to be used 
inside the light cylinder as well (see e.g.~\citealt{Toma18}). Finally, the zero 
velocity along the jet axis was reproduced in recent numerical simulation~\citep{Kousheta}.

As was already emphasised, since the results of (\ref{core})–(\ref{Bcr}) discussed above 
were obtained under the assumption of a non-zero velocity along the jet axis, it is 
important to discuss the question of whether such an internal structure of 
relativistic jets is preserved under the assumption $u(0) = 0$. The present work is 
devoted precisely to this issue. As will be shown, the seemingly fundamental difference
in the formulation of the problem does not change the key properties of the internal 
structure of relativistic jets. Moreover, relations $r_{\rm core} \approx R_{\rm L}$ 
and $B_{\rm cr} \approx \sigma_{\rm M}^{-1} B_{\rm L}$ also remain valid.

The paper is organised as follows. In Section~\ref{sec:basic_eqn}, we formulate the basic equation 
describing cylindrical cold magnetised flow. Section~\ref{sec:singular} is devoted to the 
analysis of singular points. In the problem considered here, this is the rotation 
axis, as well as the Alfv{\'e}nic  surface near the light cylinder. Finally, in
Section~\ref{sec:conclusion}, we formulate the main results of our consideration. 

\section{Basic Equations}
\label{sec:basic_eqn}

Below we use the language developed by~\citet{BHMP}: all 3D vectors 
correspond to physical quantities measured by Zero Angular Momentum Observers (which in our
case, i.e. far from the central black hole, coincides with the usual cylindrical reference 
frame). Further, it should be immediately noted that our task is not devoted to the construction
of a global solution. It is dedicated to the area far beyond the plasma generation region. 
Therefore, the region of plasma generation participates in our analysis indirectly through 
the integrals of motion, which we will try to choose in the most reasonable way.

Besides, as was shown by~\citet{BN06}, one can consider strongly collimated jet as 
a sequence of cylindrical flows. This makes it possible to explore their internal structure
by analyzing not the second-order Grad-Shafranov equation, but two first-order ordinary 
differential equations for magnetic flux $\Psi(\varpi)$ and poloidal Alfv\'enic
Mach number ${\cal M}(\varpi)$~\citep{Beskin97, BM00} 
\begin{equation}
{\cal M}^2 = \frac{4\pi\mu\eta^2}{n}.
\label{M2}
\end{equation}
Here, $n$ is the number density in the comoving reference frame and $\mu$ is 
relativistic enthalpy. Accordingly, $\eta$ is the particle-to-magnetic flux
ratio determined from relation
\begin{equation} 
n \mathbfit{u}_{\rm p} = \eta \mathbfit{B}_{\rm p},
\label{nuetaB}
\end{equation}
which is constant along magnetic field lines: $\eta = \eta(\Psi)$. 
Finally, by definition, in the cylindrical geometry
\begin{eqnarray} 
B_{z} & = & \frac{1}{2 \pi \varpi} \, \frac{{\rm d}\Psi}{{\rm d}\varpi}, \\
\label{Bp}
B_{\varphi} & = & -\frac{2I}{c\varpi}.
\nonumber
\end{eqnarray}
Here $I$ is the total electric current within the magnetic tube $\Psi =$ const. 

The first equation is the relativistic Bernoulli 
equation 
\begin{equation}
u_{\rm p}^2 = \gamma^2 - u_{\varphi}^2 - 1, 
\end{equation}
where $u_{\rm p}$ and
$u_{\varphi}$ are the poloidal and toroidal components of the 4-velocity 
$\mathbfit{u}$ respectively. It can be rewritten in the form~\citep{MHD}
\begin{eqnarray}
\frac{{\cal M}^4}{64\pi^4 \varpi^2}
\left(\frac{{\rm d}\Psi}{{\rm d}\varpi}\right)^2 =
\frac{K}{\varpi^2 A^2} - \mu^2\eta^2.
\label{ap4}
\end{eqnarray}
Here
\begin{equation}
A = 1 - \Omega_{\rm F}^2 \varpi^2/c^2-{\cal M}^2
\end{equation}
is the Alfv\'enic factor where the so-called field angular velocity $\Omega_{\rm F} = \Omega_{\rm F}(\Psi)$
is constant on the magnetic surfaces ($\Omega_{\rm F} = \Omega$ near ''the central engine''),
\begin{equation}
K= \varpi^2(e')^2(A-{\cal M}^2)+{\cal M}^4 \varpi^{2}E^2-{\cal M}^{4}L^2c^2,
\label{ap5}
\end{equation}
and by definition, 
\begin{equation}
e^{\prime}(\Psi) = E(\Psi) - \Omega_{\rm F}(\Psi)L(\Psi).
\label{eprime}
\end{equation}
Remember that Bernoulli integral $E = E(\Psi)$ and
the angular momentum flux $L = L(\Psi)$ 
\begin{eqnarray}
E(\Psi) & = & \gamma \mu \eta c^2 + \frac{\Omega_{\rm F}I}{2\pi}, 
\label{Edef} \\ 
L(\Psi) & = & \varpi u_{\varphi} \mu \eta c + \frac{I}{2\pi},    
\label{Ldef}
\end{eqnarray}
together with the angular velocity $\Omega_{\rm F}(\Psi)$ are also integrals of motion.
In this case, the current  $I$, the Lorentz factor $\gamma$, and the toroidal four-velocity $u_{\varphi}$  are expressed as follows
\begin{eqnarray}
\frac{I}{2\pi} & = & \frac{L-\Omega_{\rm F}\varpi^{2}E/c^2}
{1-\Omega_{\rm F}^{2}\varpi^{2}/c^2-{\cal M}^{2}},
\label{p33} \\
\gamma  & = & \frac{1}{\mu\eta } \, \frac{(E-\Omega_{\rm F}L)
- E{\cal M}^{2}}{1-\Omega_{\rm F}^{2}r^{2}/c^2-{\cal M}^{2}},
\label{p34} \\
u_{\varphi}  & = & \frac{1}{\mu\eta c \varpi} \, \frac{(E-\Omega_{\rm F}L)
 \Omega_{\rm F}\varpi^{2}/c^2-L{\cal M}^{2}}{1-\Omega_{\rm F}^{2}r^{2}/c^2-{\cal M}^{2}}.
\label{p35} 
\end{eqnarray}

The second equation determines the Mach number ${\cal M}$ for a cold flow 
(the sound speed $c_{\rm s} = 0$ and the relativistic enthalpy $\mu = m_{\rm p}c^2 =$ const) and is given by~\citep{MHD}:
\begin{eqnarray}
\left[\frac{(e')^2}{\mu^2\eta^2c^4}-1+\frac{\Omega_{\rm F}^2 \varpi^2}{c^2}
\right]
\frac{{\rm d}{\cal M}^2}{{\rm d} \varpi } =
\frac{{\cal M}^6L^2}{A \varpi^3 \mu^2\eta^2c^2}
\nonumber \\
+\frac{\Omega_{\rm F}^2 \varpi {\cal M}^2}{c^{2}}
\left[2 - \frac{(e')^2}{A\mu^2\eta^2c^4}\right]
+{\cal M}^2 \frac{e'}{\mu^2\eta^2c^4}\frac{{\rm d}\Psi}{{\rm d}\varpi}\frac{{\rm d}e'}
{{\rm d}\Psi} 
\label{rel-1} \\
+\frac{{\cal M}^2 \varpi^2}{2c^2}\frac{{\rm d}\Psi}{{\rm d}\varpi}
\frac{{\rm d}\Omega_{\rm F}^2}{{\rm d}\Psi}
-{\cal M}^2 \left(1-\frac{\Omega_{\rm F}^2 \varpi^2}{c^2}\right)
\frac{{\rm d}\Psi}{{\rm d} \varpi }\frac{1}{\eta}\frac{{\rm d}\eta}{{\rm d}\Psi}.
\nonumber 
\end{eqnarray}

Let us now define the integrals of motion in a convenient form. In contrast 
to the basic assumption on the finite velocity along the jet axis discussed 
earlier, we must now, following (\ref{nuetaB}),  set $\eta(\Psi) \rightarrow 0$ 
as $\Psi \rightarrow 0$. At the same time, thanks to the definitions 
(\ref{Edef})--(\ref{Ldef}), it is convenient to express the invariant 
$e^{\prime}(\Psi)$ (\ref{eprime}) in terms of the flux ratio $\eta(\Psi)$ and 
an additional function $\varepsilon(\Psi)$
\begin{equation}
(e^{\prime})^2 = \mu^2\eta^2(\Psi)c^4 - \mu^2\eta^2(\Psi)c^4\varepsilon(\Psi).
\label{eprime2}
\end{equation}
As can be seen from relations (\ref{Edef})--(\ref{Ldef}), the value of 
$\varepsilon$ vanishes for zero flow velocity. Therefore, it turns out 
to be convenient in the analysis of the problem under consideration. In
particular, the function $\varepsilon(\Psi)$ cannot have an arbitrary 
form. We clarify this issue a little later. 

Besides, following~\citep{BCKN-17, ChBP19}, we set
\begin{eqnarray}
L(\Psi) & = & \frac{\Omega_{0}\Psi}{4 \pi^2}\sqrt{1 - \frac{\Psi}{\Psi_{\rm tot}}}, 
\label{L} \\ 
\Omega_{\rm F}(\Psi) & = & \Omega_{0}\sqrt{1 - \frac{\Psi}{\Psi_{\rm tot}}}.
\label{OmegaF}
\end{eqnarray}
Such definitions ensure the closure of the longitudinal electric current within the jet.
Further, thanks to (\ref{eprime}) and (\ref{eprime2}), we have
\begin{equation}
E(\Psi) = \Omega_{\rm F}(\Psi)L(\Psi) + \mu\eta(\Psi)c^2[1 - \varepsilon(\Psi)]^{1/2}.
\label{E}
\end{equation}
Finally, due to our main assumption $\eta(0) = 0$, the fourth integral $\eta(\Psi)$, 
in the limit  $\Psi \rightarrow 0$, can be written as
\begin{equation}
\eta(\Psi) = \eta_{0}\left(\frac{\Psi}{\Psi_{\rm tot}}\right)^{\beta},
\label{eta}
\end{equation}
where $\beta > 0$. Below, for simplicity, we assume that the relation (\ref{eta}) is valid for any value of $\Psi$.

Introducing the dimensionless variables
\begin{eqnarray}
x & = & \frac{\Omega_{0}\varpi}{c},
\label{xless} \\
y & = & \frac{\Psi}{\Psi_{\rm tot}}, 
\label{yless} 
\end{eqnarray}
one can rewrite Eqns. (\ref{ap4}) and (\ref{rel-1}) as
\begin{eqnarray}
\frac{{\rm d}y}{{\rm d}x} =  \frac{\eta(y)x}{\sigma_{\rm M} |A|  {\cal M}^2} 
\left[
f(x,y)[1 - \omega^2(y)x^2 - 2 {\cal M}^2]\frac{\frac{}{}}{\frac{}{}}
\right.
\nonumber \\
  \left.
  - {\cal M}^4 \varepsilon(y) + 4 \sigma_{\rm M}{\cal M}^4\frac{\omega(y) \, l(y)}{\eta(y)} [1 - \varepsilon(y)]^{1/2} \right.
 \label{dy} \\
  \left. 
  +  4\sigma_{\rm M}^2{\cal M}^4 [\omega^2(y)x^2 - 1] \frac{l^2(y)}{x^2\eta^2(y)}
  \right]^{1/2},
\nonumber
\end{eqnarray}
\begin{eqnarray}
\frac{f(x,y)}{{\cal M}^2} \frac{{\rm d}{\cal M}^2}{{\rm d}x} = 4 \sigma_{\rm M}^2 \frac{{\cal M}^4}{A x^3}\, \frac{l^2(y)}{\eta^2(y)} + x\omega^2(y) 
\nonumber \\
- \frac{x\omega^2(y)}{A}\left[f(x,y) + {\cal M}^2\right] - \frac{1}{2} \,  \frac{{\rm d}\varepsilon(y)}{{\rm d}y} \frac{{\rm d}y}{{\rm d}x} 
\label{dM2} \\
+ \frac{1}{2} \, x^2 \frac{{\rm d}\omega^2(y)}{{\rm d}y} \frac{{\rm d}y}{{\rm d}x}
+ \frac{1}{2} \,   \frac{f(x,y)}{\eta^2(y)} \, \frac{{\rm d}\eta^2(y)}{{\rm d}y} \,\frac{{\rm d}y}{{\rm d}x}.
\nonumber 
\end{eqnarray}
Here
\begin{equation}
\sigma_{\rm M} = \frac{\Omega_{0}^2\Psi_{\rm tot}}{8 \pi^2 \mu \eta_{0}c^2}
\label{sigmaM}
\end{equation}
is the Michel magnetisation parameter  already mentioned above, $\eta(\Psi) = \eta_{0}\eta(y)$, and now
\begin{equation}
A = 1 - \omega^2(y) x^2 - {\cal M}^2.
\label{AA}
\end{equation}
 Further, we introduce new important function
\begin{equation}
f(x,y) = \omega^2(y)x^2 - \varepsilon(y).
\label{f}
\end{equation}
Finally, despite the fact that according to (\ref{L}), (\ref{OmegaF}) 
and (\ref{eta}), we have $l(y) = y(1 - y)^{1/2}$, $\omega(y) = (1 - y)^{1/2}$, 
and $\eta(y) = y^{\beta}$, we have kept their literal expressions in Eqns. 
(\ref{dy}) and (\ref{dM2}).

\section{Singular points}
\label{sec:singular}
\subsection{Rotation axis}

Before integrating Eqns. (\ref{dy})--(\ref{dM2}), let us discuss their behaviour 
for $x \rightarrow 0$. This helps us with numerical integration as well. 
Below we assume that poloidal magnetic field $B_{z}$ and the number density 
$n$ are finite at the rotation axis. Then, due to definition (\ref{M2}),  
${\cal M}^2 \rightarrow 0$  if $\eta \rightarrow 0$.
Storing now only the leading terms (and grouping the similar ones), we obtain
\begin{eqnarray}.
\frac{{\rm d}y}{{\rm d}x} =  \frac{\eta(y)x}{\sigma_{\rm M}{\cal M}^2} 
%\left[x^2 - \varepsilon(y)\right]^{1/2},
f^{1/2},
\label{dy0}
\end{eqnarray}
\begin{eqnarray}
\frac{f^{3/2}\eta(y)}{{\cal M}^2} \frac{{\rm d}}{{\rm d}x}
\left[\frac{{\cal M}^2}{f^{1/2}\eta(y)}\right] + (f + {\cal M}^2)x =
4 \, \sigma_{\rm M}^2 \frac{{\cal M}^4y^2}{x^3\eta^2(y)}.
\label{dM20}
\end{eqnarray}

As one can see, the function $f(x,y)$ plays the primary role in determining 
the behaviour of the solution near the rotation axis, and thus, the function 
$\varepsilon(y)$ should be introduced. In order to understand the functional 
form of $\varepsilon(y)$ for our problem statement, let us suppose that the
magnetic field is regular at $x \rightarrow 0$. In this case, it is convenient 
to introduce the dimensionless magnetic field
\begin{equation}
b = \frac{B_{z}}{B_{\rm L}},
\label{b}
\end{equation}
where $B_{\rm L}$ is the magnetic field on the light cylinder near the origin 
and can be determined from the condition $\Psi_{\rm tot} = \pi R_{\rm L}^2 B_{\rm L}$. It gives
\begin{equation}
b(x) = \frac{1}{2x}\, \frac{{\rm d}y}{{\rm d}x}.
\label{bx}
\end{equation}
In particular, denoting $b_{0}=b(0)$, we get for $x \rightarrow 0$
\begin{equation}
y(x) \approx b_{0}x^2.
\label{bxx}
\end{equation}
It is clear that in what follows we will be interested in the case $b_{0} \ll 1$,
because the light cylinder must contain only a small part of the total magnetic 
flux as the size of the jet is much larger than the light cylinder.

Further, according to (\ref{Edef})--(\ref{Ldef}), we have for $v \ll c$, 
\begin{equation}
\frac{e^{\prime}(\Psi)}{\mu\eta(\Psi)c^2} = \gamma - \frac{\Omega_{\rm F}\varpi}{c}u_{\varphi} 
= 1 + \frac{1}{2} \, \frac{v_{\rm p}^2}{c^2} + \frac{1}{2} \, \frac{v_{\varphi}^2}{c^2} - \frac{\Omega_{\rm F}\varpi}{c}\, \frac{v_{\varphi}}{c}.
\label{e1}
\end{equation}
Comparing this expression with the definition (\ref{eprime2}),
we obtain 
\begin{equation}
\varepsilon(y) = 2\frac{\Omega_{\rm F}\varpi}{c}\, \frac{v_{\varphi}}{c} - \frac{v_{\rm p}^2}{c^2} - \frac{v_{\varphi}^2}{c^2},
\label{e2}
\end{equation}
and thus, according to (\ref{f}), we have
\begin{equation}
f(x, y) = \frac{(v_{\varphi} - \Omega_{\rm F}\varpi)^2}{c^2} + \frac{v_{\rm p}^2}{c^2}.
\label{fxyv}
\end{equation}
However, as is well-known (see, e.g.,~\citealt{MHD}), relation (\ref{p35}) 
gives  $v_{\varphi}\rightarrow \Omega_{\rm F}\varpi$ for $\varpi \rightarrow 0$.
Thus, $f(x,y) \rightarrow v_{\rm p}^2/c^2$ as $x \rightarrow 0$. Using now definitions 
(\ref{M2}) and (\ref{nuetaB}), we return to relation (\ref{dy0}). 

This result is certainly an important confirmation of the consistency 
of our approach. Moreover, it allows us to use relation (\ref{f}) as a definition 
of $\varepsilon(y)$ for $y \rightarrow 0$. Together with (\ref{dy0}), it gives
\begin{equation}
\varepsilon(y)  = \frac{y}{b_{0}} - 4 \, \eta^2(y) \, b_{0}^2 \, {\cal M}_0^4 \, \sigma_{\rm M}^2.
\label{varydef}
\end{equation}
Here we introduce one more parameter
\begin{equation}
{\cal M}_0^2 = \frac{4 \pi \mu \eta_{0}^2}{n_{0}},
\label{M02deff}
\end{equation}
specifying the particle number density on the rotation axis $n_{0} = n(0)$.

Relation (\ref{varydef}) immediately allows us to make two important conclusions. 
Indeed, since $\varepsilon(y)$ is only a function of $y$, it cannot depend on such parameters 
as the magnetic field $b_{0}$ and the number density $n_{0}$ on any particular slice.
This becomes possible only if the conditions  
\begin{equation}
\eta(y) = y^{1/2}
\label{etaeta}
\end{equation}
and 
\begin{equation}
\frac{1}{b_{0}} -  4 \, b_{0}^2 \, {\cal M}_0^4 \, \sigma_{\rm M}^2 = {\cal C},
\label{b0n0}
\end{equation}
where ${\cal C} =$ const., are met. The first of them fixes the behaviour 
of the function $\eta(y)$ for $y \rightarrow 0$. As was already stressed, in what follows 
we assume that condition (\ref{etaeta}) is valid for all values of $y$. As for relation 
(\ref{b0n0}), we must now consider it as a connection between the magnetic field 
$b_{0}$ and the number density $n_{0}$ on the jet axis. Further,
for estimates we can set ${\cal C} = 0$, so that 
$2 \, b_{0} \, {\cal M}_0^2 \, \sigma_{\rm M} \approx b_{0}^{-1/2} \gg 1$.

Returning now to Eqns. (\ref{dy0})--(\ref{dM20}) in the limit $x \rightarrow 0$, let us rewrite them in the form
\begin{equation}
{\tilde\beta}(x) =  \frac{1}{2 \, b_{0}\sigma_{\rm M}} \, \frac{\eta(y)f^{1/2}}{{\cal M}^2(x)},
\label{db1}
\end{equation}
\begin{equation}
-\frac{1}{{\tilde\beta}(x)} \frac{{\rm d}{\tilde\beta}(x)}{{\rm d}x}
 + \left(1 + \frac{\zeta}{{\tilde\beta}^2(x)}\right) x =
\frac{x}{{\tilde\beta}^2(x)}.
\label{db2}
\end{equation}
Here
\begin{equation}
\zeta =  \frac{1}{4 \, b_{0}^2{\cal M}_{0}^2 \sigma_{\rm M}^2}
= \frac{4 \pi \mu n_{0}}{B_{0}^2} \ll 1,
\label{zeta}
\end{equation}
and ${\tilde\beta}(x) = b(x)/b_{0}$ so that $\beta(0) = 1$.
As one can see, Eqn. (\ref{db2}) is regular at $x \rightarrow 0$. It describes 
the change of the magnetic field. Actually, it depends on only one parameter
$\zeta$ (\ref{zeta}), which is small due to condition $b_{0} \ll  1$. This 
confirms our assumption that the magnetic field remains finite at $x \rightarrow 0$.

\begin{figure} 
\begin{center}
\includegraphics[width=0.9\columnwidth]{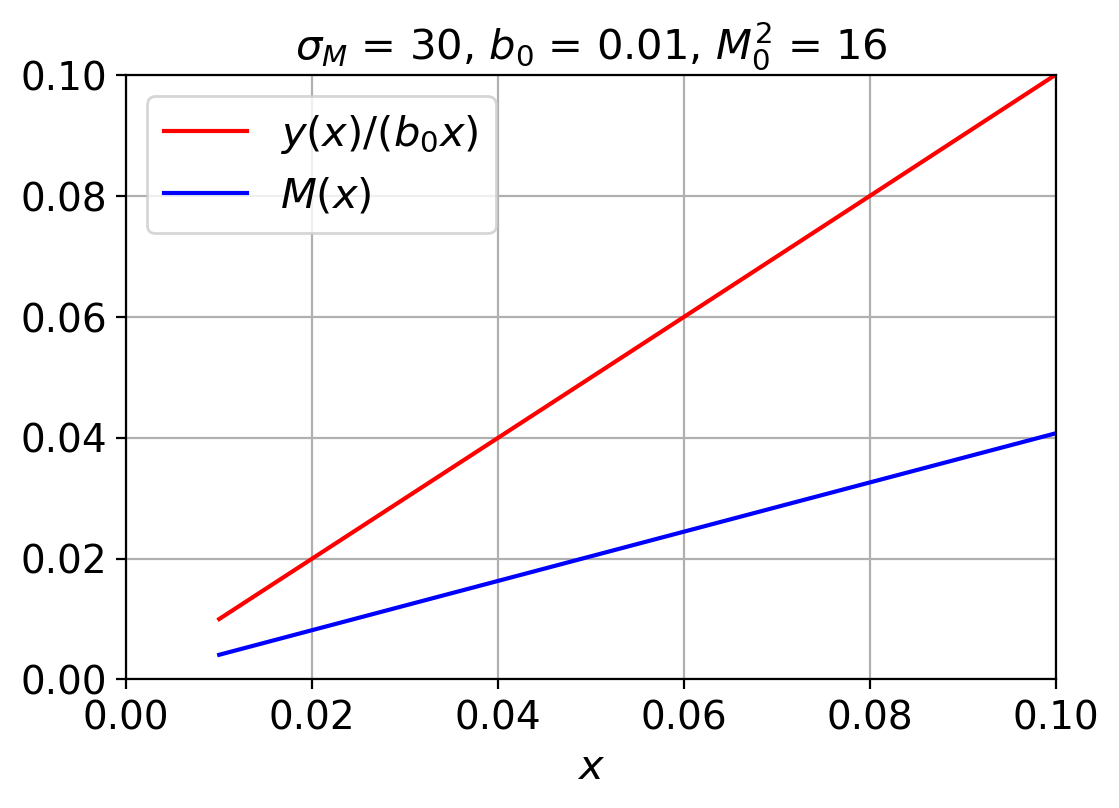}
  \end{center}
  \caption{Change in $y(x)/(b_{0}x)$ and ${\cal M}(x)$ for small $x$, 
  obtained as a solution of exact Eqns. (\ref{dy})--(\ref{dM2}) using 
  boundary conditions $y(x_0) = b_{0}x_{0}^2$ and ${\cal M}^2(x_0) = x_{0}^2$ for $x_{0} = 0.01$. }
\label{fig1}
\end{figure}

As for Eqn. (\ref{db1}), it can be now used to determine ${\cal M}^2(x)$ in the 
limit $x \rightarrow 0$. It finally gives
\begin{equation}
{\cal M}^2(x) \approx b_{0} {\cal M}_0^2 \, x^2.
\label{Mxx}
\end{equation}
In Figure~\ref{fig1}, we show the change in $y(x)/(b_{0}x) \approx x$ and 
${\cal M}(x) \propto x$ for small $x$, obtained as an exact solution of 
Eqns. (\ref{dy})--(\ref{dM2}) using boundary conditions $y(x_0) = b_{0}x_{0}^2$ 
and ${\cal M}^2(x_0) = b_{0} {\cal M}_0^2 x_{0}^2$ for $x_{0} = 0.01$. 
As one can see, the exact solution is in full agreement with the analytical estimates
(\ref{bxx}) and (\ref{Mxx}).

\subsection{Alfv{\'e}nic surface}

Before proceeding to a discussion of the general structure of a poloidal 
magnetic field outside the light cylinder, it is necessary to discuss 
the critical conditions on the Alfv{\'e}nic surface $A = 0$. As for the fast 
magnetosonic surface, there is no singularity on it in the cylindrical 
geometry considered here~\citep{MHD}. This well-known effect is similar 
to the shift of the singularity into the modified fast magnetosonic
surface in the self-similar~\citet{1982MNRAS.199..883B} solution. For
cylindrical geometry, this singularity shifts to infinity.

As for the critical condition on the Alfv{\'e}nic surface, it is more 
convenient to find it from the numerator of relation (\ref{p34}):
\begin{equation}
e^{\prime}(\Psi_{\rm A}) = E(\Psi_{\rm A}){\cal M}^2(r_{\rm A}).
\label{PsiA}
\end{equation}
Here all the quantities are to be taken at the Alfv{\'e}nic point,
so that $\Psi_{\rm A} = \Psi(r_{\rm A})$. It is easy to check that,
in this case, the regularity conditions in relations (\ref{p33}) and
(\ref{p35}), as well as in our basic equations (\ref{dy})--(\ref{dM2}), 
are automatically fulfilled.

Note now that for the strongly magnetised flow (${\cal M}^2 \ll 1$) under discussion, 
the Alfv{\'e}nic surface is located near the light cylinder: $r_{\rm A} \approx R_{\rm L}$
(i.e. $x_{\rm A} \approx 1$). Using the dimensionless variables (\ref{xless})--(\ref{yless}) 
introduced above, one can rewrite the critical condition (\ref{PsiA}) as
\begin{equation}
2 \, \sigma_{\rm M} \, {\cal M}_0^2 \, y \, \eta(y) = 1.
\label{xA}
\end{equation}
Taking into account relations (\ref{bxx}) and (\ref{etaeta}) as well as under condition
$x_{\rm A} \approx 1$, we finally obtain
\begin{equation}
4 \, \sigma_{\rm M}^2 \, {\cal M}_0^4  \approx b_{0}^{-3}.
\label{Alfv}
\end{equation}
As we see, condition (\ref{Alfv}) is in accordance with relation (\ref{b0n0}) for ${\cal C} \approx 0$.
Therefore, we will not dwell on the problem of passing the critical  surface in detail and will immediately proceed 
to the analysis of the solution for $r > R_{\rm L}$ (or $x > 1$).

\section{Discussion and conclusion}
\label{sec:conclusion}

\begin{figure} 
\begin{center}
\includegraphics[width=0.9\columnwidth]{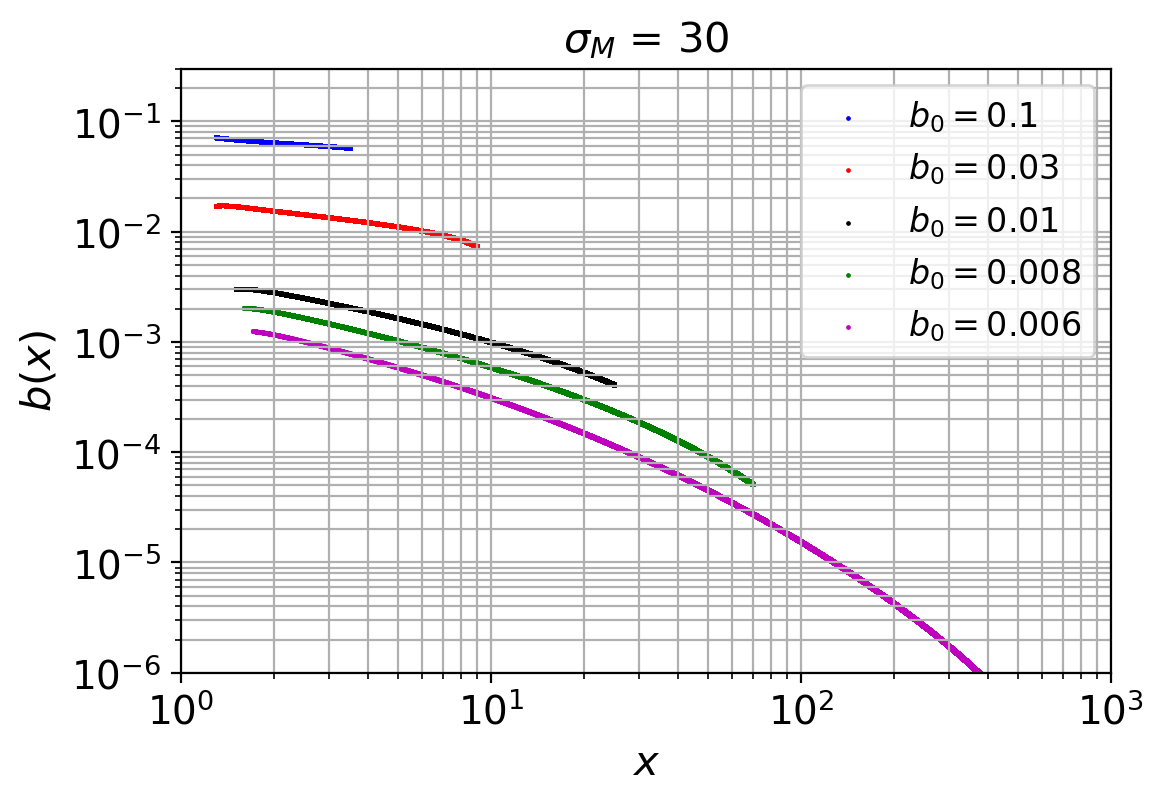}
  \end{center}
  \caption{Dimensionless magnetic field $b(x)$ obtained from solutions 
    of general equations (\ref{dy})--(\ref{dM2}).
   The jet size $x_{\rm jet} = r_{\rm jet}/R_{\rm L}$ is determined
   from the condition $\Psi(r_{\rm jet}) = \Psi_{\rm tot}$.
  }
\label{fig2}
\end{figure}

In Figure~\ref{fig2}, we show solutions of general equations (\ref{dy})--(\ref{dM2})
for dimensionless magnetic field $b(x)$. We carry out the integration from the region of a singular point with 
boundary conditions corresponding to the asymptotic solutions (\ref{bxx}) and (\ref{Mxx}) for
$x = 1$. For this reason, the main control parameter is the magnetic field $b_{0}$ on the jet axis. 
The jet size $r_{\rm jet}$ is determined from the condition $\Psi(r_{\rm jet}) = \Psi_{\rm tot}$. 

\begin{figure} 
\begin{center}
\includegraphics[width=0.9\columnwidth]{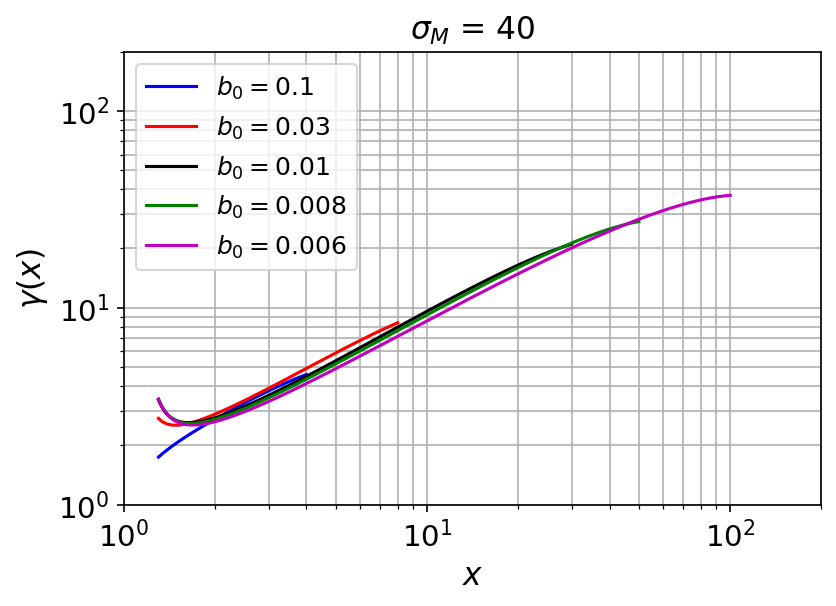}
  \end{center}
  \caption{Lorentz-factor of particles $\gamma(x)$ as a function of distance from the axis $x$ 
  at different distances from ''the central engine'' outside the light cylinder. The downward 
  bend at large $x$ is associated with a rather small value of $\sigma_{\rm M}$, which 
  determines the maximum possible value of $\gamma$.
  }
\label{fig3}
\end{figure}

As one can see, despite the fact that the velocity at the axis vanishes, in general, 
there is complete qualitative agreement with the results obtained under the assumption of 
a finite flow velocity near the axis (see, e.g.,~\citealt{BN09, Lyu09}). The poloidal 
magnetic field $B_{z}$ remains practically constant within the light cylinder. As for 
the structure of the magnetic field outside the light cylinder, it depends on the magnetic field 
$b_{0}$ on the jet axis. For sufficiently large values of $b_{0}$, longitudinal 
magnetic field remains essentially uniform ($B_{\rm z} \approx $ const). But for 
small values of $b_{0}$, a central core begins to form near the jet axis, the size 
of which, however, does not tend to zero, as might be expected according to (\ref{core}).
In all cases, its size remains on the order of the radius of the light cylinder:
\begin{equation}
r_{\rm core} \approx R_{\rm L},
\label{corecorr}
\end{equation}

Additionally, there is a quantitative agreement if the expression (\ref{Bcr}) is corrected to
\begin{equation}
B_{\rm cr} \approx \frac{B_{\rm L}}{\sigma_{\rm M}}.
\label{Bcrcorr}
\end{equation}
For $\sigma_{\rm M} = 30$ shown in Figure~\ref{fig2}, expression (\ref{Bcrcorr}) results 
in $b_{0} = 0.03$ for the critical magnetic field. As one can see, this is exactly what takes place.
Finally, as shown in Figure~\ref{fig3}, the universal asymptotic behavior $\gamma \approx x$ is also 
reproduced with good accuracy outside the light cylinder.

\begin{figure} 
\begin{center}
\includegraphics[width=0.9\columnwidth]{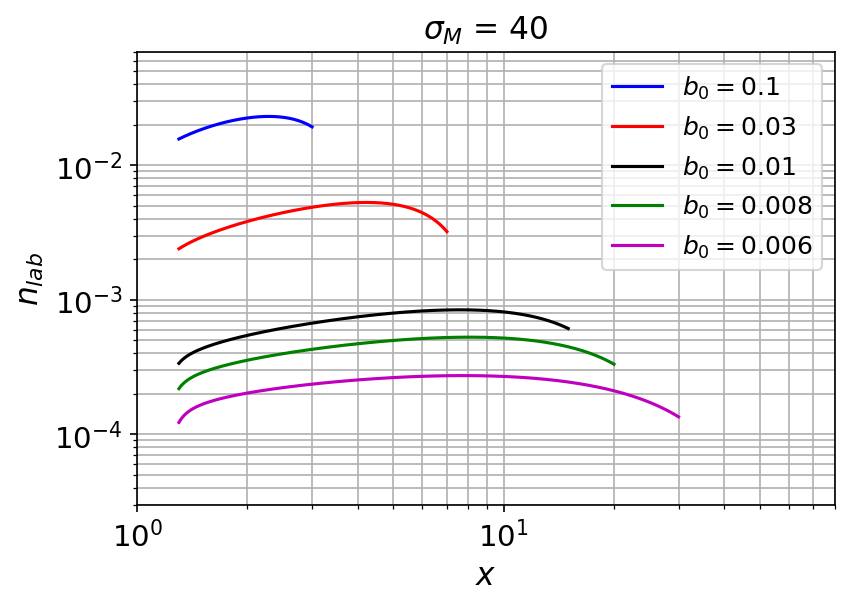}
  \end{center}
  \caption{Particle number density $n_{\rm lab} = n \gamma$ in the laboratory reference frame
  as a function of distance from the axis $x$  at different distances from ''the central engine'' 
  outside the light cylinder.
  }
\label{fig4}
\end{figure}

On the other hand, we found one significant difference between the commonly considered case  
$\eta(y) \approx $ const and the case $\eta(y) = y^{1/2}$ considered in this paper. As shown 
in Figure~\ref{fig4}, particle number density $n_{\rm lab} = n \gamma$ in the laboratory reference 
frame remains almost constant outside the central core. This difference, however, can easily be 
explained. 

Indeed, according to definition (\ref{M2}), the number density in the comoving reference frame can 
be written as $n = 4 \pi \mu \eta^2/{\cal M}^2$. Further, far from the light cylinder 
($\Omega_{\rm F}^2\varpi^2/c^2 \gg 1$), but in the region of a strongly magnetized flow 
(${\cal M}^2 \ll \Omega_{\rm F}^2\varpi^2/c^2$), Lorentz factor $\gamma$ according to (\ref{p34}) 
has the form
\begin{equation}
\gamma \approx \frac{{\cal M}^2 E c^2}{\mu\eta\Omega_{\rm F}^2\varpi^2}.
\end{equation}
Using now relations (\ref{L})--(\ref{E}) to determine Bernoully integral $E$, we finally obtain
\begin{equation}
n_{\rm lab} \approx \frac{\eta \Psi}{\pi \varpi^2}.
\end{equation}
As a result, at a constant $\eta$ and in the region of existence of the central core, when magnetic
flux $\Psi$ grows slowly than $\varpi^2$, the number density  $n_{\rm lab}$ is to decrease with 
increasing distance $\varpi$ from the axis. On the other hand, in the case $\eta = y^{1/2}$, 
depending on the behavior of the solution $\Psi = \Psi(x)$, both an increase and a decrease in 
the number density $n_{\rm lab}$ with distance $x$ from the axis are possible.
Here, however, it should be noted that such behavior takes place only if the
relation $\eta = y^{1/2}$ remains valid up to the jet boundary. If this dependence takes place only
at $x \rightarrow 0$, and at $x \sim 1$ we have $\eta \approx$ const, then the number density $n_{\rm lab}$
is to decrease with the distance from the axis.

\begin{figure} 
\begin{center}
\includegraphics[width=1.0\columnwidth]{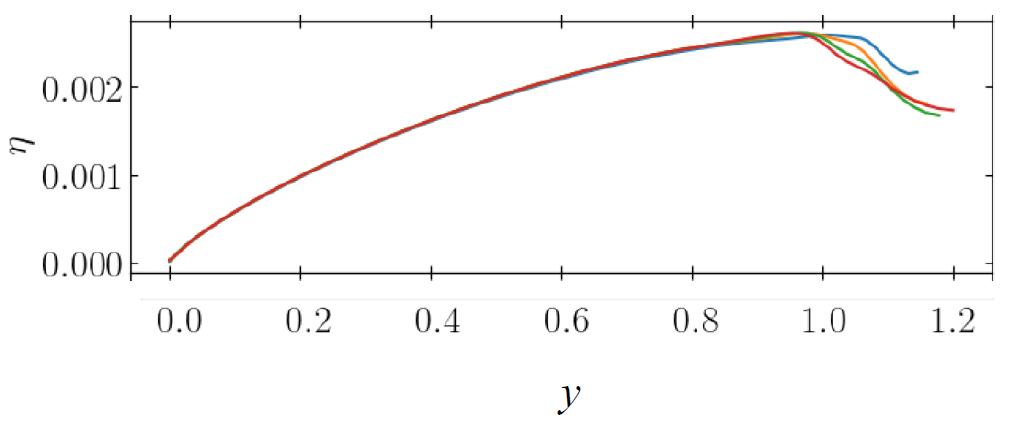}
  \end{center}
  \caption{Function $\eta(y)$ reproduced from the results of a numerical simulation 
  carried out by~\citet{Kousheta}. Different curves correspond to different distances 
  from ''the central engine'', confirming that $\eta(y)$ is indeed an integral of motion.
  }
\label{fig5}
\end{figure}

\begin{figure} 
\begin{center}
\includegraphics[width=0.9\columnwidth]{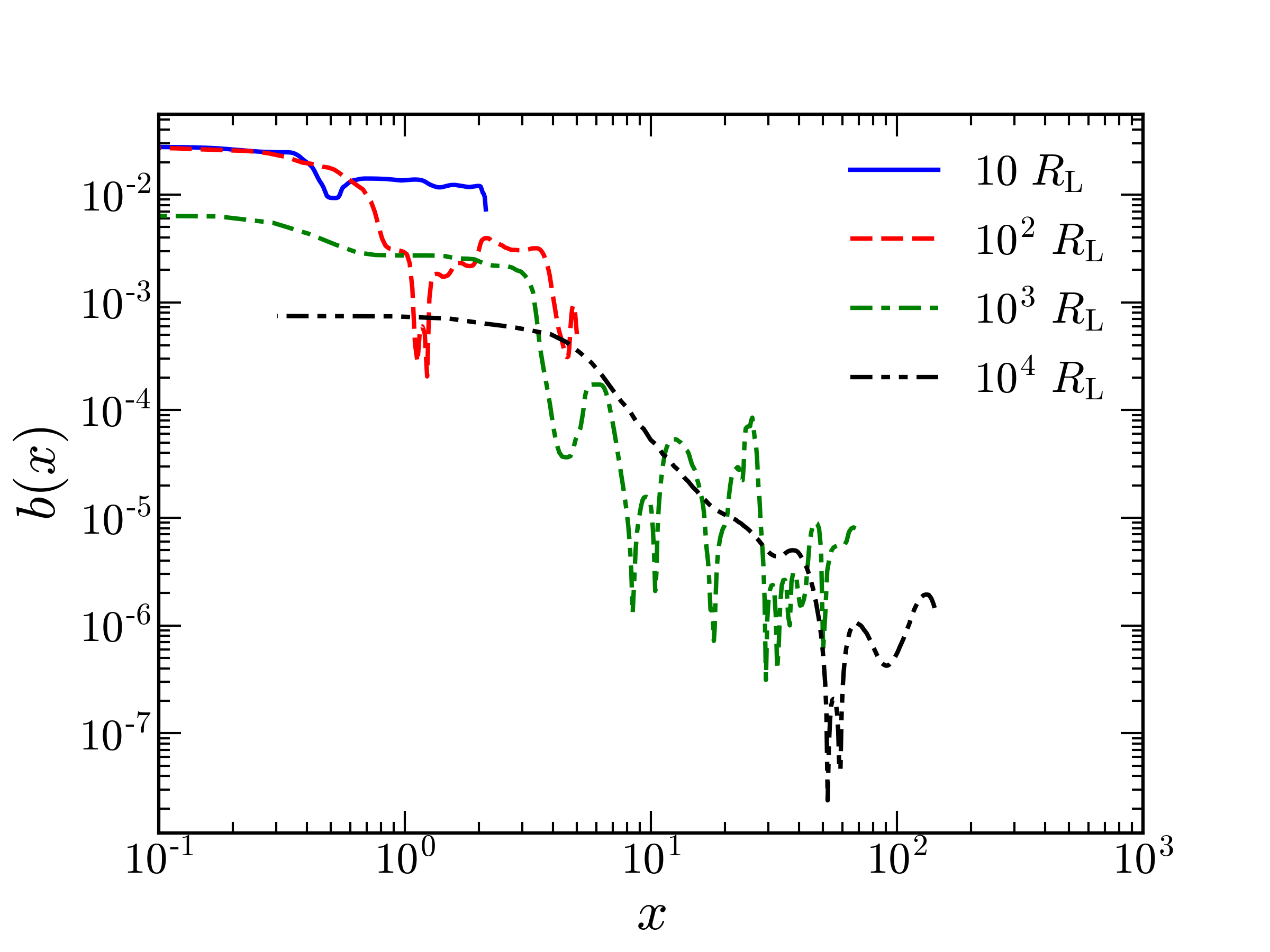}
  \end{center}
  \caption{Dependence of the dimensionless poloidal magnetic field $b(x)$ on the 
  distance to the axis $x = \varpi/R_{\rm L}$ at different distances from the 
  origin obtained by numerical simulation carried out by~\citet{Kousheta}.
  The non-uniform behaviour of $b(x)$ at large $x$ occurs to the jet boundary instabilities.}
\label{fig6}
\end{figure}

Moreover, our analytical results are in excellent agreement with the above-mentioned results 
of numerical simulations of~\citet{Kousheta}. First, Figure~\ref{fig5} shows that jets in numerical 
simulations exhibit the dependence $\eta(y) \propto y^{1/2}$ (\ref{etaeta}), surprisingly matching 
the relation (\ref{etaeta}). Here, different curves correspond to different distances from 
''the central engine'', confirming that $\eta(y)$ is indeed an integral of motion. We emphasise 
that the value of the integral $\eta(\Psi)$, like all other integrals of motion, was not set 
initially, as is done in analytical calculations, but emerged self-consistently as a result 
of evolving a time-dependent numerical simulation. Second, as shown in Figure~\ref{fig6}, the 
dependence of the dimensionless poloidal magnetic field $b(x)$ on the dimensionless distance
to the axis $x = \varpi/R_{\rm L}$ at different distances from the origin in the simulation 
also well reproduces the structure of the poloidal field shown in Figure~\ref{fig2}.
As far as the number density distribution is concerned, it is determined by the magnetic field 
strength in the form of the so-called density floors~\citep{Porth2019}, which does not allow us to 
determine it with sufficient accuracy. Therefore, we do not present here the results of numerical 
simulation concerning the quantity $n_{\rm lab}$.

Thus, we can state with confidence that the appearance of a central core at sufficiently 
large distances from ``the central engine'' does not depend on the plasma flow velocity near 
the jet axis. In all cases, at a sufficiently low ambient pressure, a dense
core appears near the axis, the radius of which is close to the size of the light cylinder.
Outside the central core, both the poloidal magnetic field and the plasma number density decrease 
with a power-law behaviour.

Finally, our results hold important implications for the jet structure and velocities 
at distances far from the black hole, relevant for interpreting observed jet morphologies 
and widths, as well as the transverse jet velocity stratification measured in AGN 
jets~\citep[as was seen by][for the M87 jet]{Mertens, add1}. Indeed, the presence of a 
central core region and low velocity region at the jet axis was also seen in global 
semianalytical work~\citep{add2,add3}. As we show, once the central core forms atdistances 
$z>z_{cr}$ from the black hole, the poloidal magnetic field in the jet becomes of the order 
of $B_{cr}$, the jet becomes susceptible to magnetic pinch and kink instabilities. This
result is verified in 2D and 3D numerical simulations~\citep[][]{2016MNRAS.456.1739B,Kousheta}.
Thus, we suggest that when a central core appears, the observed width of the jet will be 
determined precisely by the magnetically dominated inner jet region, and not by the 
geometric width of the jet.

\section*{Data availability}
The data underlying this work will be shared on reasonable request to the corresponding author.

\section{Acknowledgements}

We thank Anna Chashkina and Alexander Tchekhovskoy for useful discussions. This work was partially 
supported by the National Research Center Kurchatov Institute (Order No. 85 dated 03.20.23). 
KC is supported by the Black Hole Initiative at Harvard University, which is funded by grants 
from the Gordon and Betty Moore Foundation, John Templeton Foundation and the Black Hole PIRE 
program (NSF grant OISE-1743747).

\bibliography{zero}
\bibliographystyle{mnras}

\end{document}